\documentclass[10pt, conference, letterpaper]{IEEEtran}
\IEEEoverridecommandlockouts
\usepackage{booktabs} 
\usepackage{amsmath}
\usepackage[english]{babel}
\usepackage[latin1]{inputenc}
\usepackage{amssymb}
\usepackage[ruled]{algorithm2e}
\usepackage{algpseudocode}
\usepackage{bm}
\usepackage{amsmath}
\usepackage{graphicx}
\usepackage{epsfig}
\usepackage{url}
\usepackage{psfrag}
\usepackage{balance}
\usepackage{pstricks}
\usepackage{times}
\usepackage{array}
\usepackage{subfigure}
\usepackage{float}
\usepackage{multirow}
\usepackage{mathtools}

\algnewcommand\algorithmicforeach{\textbf{for each}}

\begin{document}

\title{A Fast Sketch Method for Mining User Similarities\\over Fully Dynamic Graph Streams
\thanks{\textsuperscript{*}Pinghui Wang is the corresponding author.}}
\author{
      \fontsize{11}{11}\selectfont
      Peng Jia$^{1}$, Pinghui Wang$^{2,1}$, Jing Tao$^{1,2,3}$, and Xiaohong Guan$^{1,2,4}$\\
      $^{1}$MOE Key Laboratory for Intelligent Networks and Network Security, Xi'an Jiaotong University, China\\
      $^{2}$Shenzhen Research Institute, Xi'an Jiaotong University, Shenzhen, China\\
      $^{3}$Zhejiang Research Institute, Xi'an Jiaotong University, Hangzhou, China\\
      $^{4}$Department of Automation and NLIST Lab, Tsinghua University, Beijing, China\\
      Email: \{pengjia, phwang, jtao, xhguan\}@sei.xjtu.edu.cn
}

\maketitle
\begin{abstract}
Many real-world networks such as Twitter and YouTube are given as fully dynamic graph streams represented as sequences of
edge insertions and deletions. (e.g., users can subscribe and unsubscribe to channels on YouTube).
Existing similarity estimation methods such as MinHash and OPH are customized to static graphs.
We observe that they are indeed sampling methods and  exhibit a sampling bias when applied to fully dynamic graph streams,
which results in large estimation errors.
To solve this challenge, we develop a fast and accurate sketch method VOS.
VOS processes each edge in the graph stream of interest with small time complexity $O(1)$
and uses small memory space to build a compact sketch of the dynamic graph stream over time.
Based on the sketch built on-the-fly, we develop a method to estimate user similarities over time.
We conduct extensive experiments and the experimental results demonstrate the efficiency and efficacy of our method.
\end{abstract}

\section{Introduction} \label{sec:introduction}
Many real-world network systems such as online social networks (OSNs) and mobile phone networks are given as graph streams represented as sequences of edges over time,
where entities are modeled as nodes and entity relations are modeled as edges.
Estimating the similarities of users in large graph streams has been successfully used for applications such as duplicate detection~\cite{Xia2011silo} and collaborative filtering~\cite{Guo2015trustsvd}.
However, the graph streams studied in all these works only consist of edge insertions.
In practice, real-world networks contain not only edge insertions but also deletions.
For example, users on OSNs such as Twitter and Pinterest can follow other users, and can also unfollow users that they followed previously;
users on YouTube can subscribe to interested channels, and can also unsubscribe from channels that they subscribed previously.

For similarity estimation, MinHash~\cite{BroderSTOC2000} is a popular sketch method for approximately computing the Jaccard coefficient similarity,
which builds a sketch of $k$ registers with $k$ distinct hash functions for each user and updates the minimum hash values of its subscribed items for each register.
To reduce the memory usage of MinHash,
\cite{PingWWW2010,MitzenmacherWWW14} develop methods \emph{b-bit minwise hashing} and \emph{odd sketch},
and the basic idea behind them is to use probabilistic methods such as sampling and sketching to build a compact digest for each user's MinHash sketch.
b-bit minwise hashing, odd sketch, and MinHash update each item with a high time complexity $O(k)$.
To solve this problem, Li et al.~\cite{Linips2012} further develop a method OPH and use only one hash functions to reduce the time complexity of updating each item from $O(k)$ to $O(1)$.
Also there are many other works based on OPH such as~\cite{ShrivastavaUAI2014,ShrivastavaICML2014,ShrivastavaICML2017} to improve its estimation accuracy.
In detail, they fill empty registers generated from OPH by ``rotation" with the value of the closest non-empty registers towards right~\cite{ShrivastavaICML2014}, left or right with probability $\frac{1}{2}$~\cite{ShrivastavaUAI2014}, or based on tailored 2-universal hashing~\cite{ShrivastavaICML2017}.
\cite{Ioffe2010improved,ShrivastavaNIPS2016,WuICDM2016,WuWWW2017} develop a fast method to estimate the Jaccard coefficient between
weighted vectors,
where the general Jaccard coefficient between two positive real value vectors $\vec x\!=\!(x_1, x_2, \ldots, x_p)$ and $\vec y\!=\!(y_1, y_2, \ldots, y_p)$ is defined as $J(\vec x, \vec y)\!=\! \frac{\sum_{1\le j\le p}\min (x_j, y_j)}{\sum_{1\le j\le p}\max (x_j, y_j)}$.
Unfortunately, all these methods indeed are sampling methods customized for statistic datasets,
but fail to uniformly sample edges from fully dynamic graph streams including item-subscriptions and item-unsubscriptions,
which results in large estimation errors.


To solve the above challenges, we develop a fast and accurate sketch method VOS (virtual odd sketch) for estimating the similarities of users occurred in fully dynamic graph streams.
VOS processes each edge with small time complexity $O(1)$ and uses small memory space to build a compact sketch of the graph stream over time.
For each user, we build an odd sketch of its subscribed items on the fly,
which is a binary sketch of $k$ bits and embeds each subscribed item with xor (i.e., exclusive-or) operations.
In graph streams, it is wasteful to assign a large $k$ for each user to achieve reasonable estimation accuracy, especially for users with few subscribed items.
Thus instead of directly keeping the odd sketch in memory,
we store the sketch in $k$ bits randomly selected from a shared bit array to reduce the memory usage.
Based on the built virtual sketch, we develop a novel method to accurately estimate user similarities,
and provide a theoretical proof for the estimation accuracy.
We conduct extensive experiments on a variety of real-world graphs,
and experimental results show that our method VOS is more accurate than state-of-the-art methods.
\section{Problem Formulation} \label{sec:problem}
In this paper, we focus on bipartite graphs,
while our method can be easily extended to regular graphs.
Let $U$ and $I$ denote the set of users and items respectively.
Let $\Pi\!=\!e^{(1)} e^{(2)} \cdots e^{(t)} \cdots$ denote the graph stream of interest,
where $e^{(t)}\!=\!(u^{(t)}, i^{(t)}, a^{(t)})$ is the element (or, edge) of $\Pi$ occurred at discrete time $t> 0$,
$u^{(t)}\in U$, $i^{(t)}\in I$, and $a^{(t)}\in \{``+", ``-"\}$ are the $t^\text{th}$ element's user, item, and action (i.e., subscription and unsubscription).
Let $S_u^{(t)}$ be the set of items subscribed by user $u$ at the end of time $t$, $S_u^{(0)}=\emptyset$.
Similar to~\cite{GemullaVLDBJ2008,StefaniKDD16},
we restrict attention to ``\emph{\textbf{feasible}}" fully dynamic graph steams.
In detail, if item $i$ is in (resp. not in) the item set $S_u^{(t-1)}$ of user $u$,
then element $(u, i, ``+")$ (resp. element $(u, i, ``-")$) cannot occur in stream $\Pi$ at time $t$.
Let $s_{u,v}^{(t)}$ denote the number of common items that users $u$ and $v$ subscribe to at time $t$
and is computed as $s_{u,v}^{(t)}\!=\!|S_u^{(t)}\cap S_v^{(t)}|\!=\!\frac{J(S_u^{(t)}, S_v^{(t)}) (|S_u^{(t)}| + |S_v^{(t)}|)}{J(S_u^{(t)}, S_v^{(t)})+1},$
where $|S|$ refers to the cardinality of a set $S$.
One can use a counter to easily keep tracking of the number of items (i.e., $|S_u^{(t)}|$) subscribed by each user $u$ over time $t$.
Besides, another popular similarity measure the Jaccard coefficient $J(S_u^{(t)}, S_v^{(t)})\!=\!\frac{|S_u^{(t)}\cap S_v^{(t)}|}{|S_u^{(t)}\cup S_v^{(t)}|}\!=\!\frac{s_{u,v}^{(t)}}{|S_u^{(t)}| + |S_v^{(t)}| - s_{u,v}^{(t)}}$
can be easily computed from $s_{u,v}^{(t)}$, and vice versa.
In this paper, we aim to develop a fast and accurate method to estimate $s_{u,v}^{(t)}$ and $J(S_u^{(t)}, S_v^{(t)})$ for any two users $u$ and $v$ over time.

\section{Shortcomings of Existing Methods}\label{sec:preliminaries}
For any two sets $S_1$ and $S_2$, 
MinHash~\cite{BroderSTOC2000} applies $k$ independent hash functions $h_1, \ldots, h_k$ to obtain an accurate estimation of $J(S_1, S_2)$,
where any hash function $h_j, 1 \leq j \leq k$ can be described as a random permutation from $I$ to $I$ itself.
For a set $S \subset I$, let $h_j^*(S)$ denote the minimum hash value of items in $S$ with respect to hash function $h_j$,
i.e., $h_j^*(S)\!=\!\min_{i\in S} h_j(i).$
Therefore, MinHash computes $h_1^*(S_1)$, $\ldots$, $h_k^*(S_1)$ and $h_1^*(S_2)$, $\ldots$, $h_k^*(S_2)$,
and then estimates $J(S_1, S_2)$ as $J(S_1, S_2)\!=\!\frac{\sum_{j=1}^k \mathbf{1}(h_j^*(S_1)  = h_j^*(S_2))}{k},$
where $\mathbf{1}(\mathbb{P})$ is an indicator function that equals 1 when predicate $\mathbb{P}$ is true and 0 otherwise.
Actually, the MinHash sketch of a set $S$ can be viewed as a vector of $k$ items sampled with replacement from $S$ using $k$ hash functions respectively.
Denote by $\phi_j(S)$  the item in $S$ with the minimum hash value with respect to hash function $h_j$,
i.e., $\phi_j(S)\!=\!\arg \min_{i\in S} h_j(i).$
Because hash function $h_j$ maps items in $I$ into distinct integers, i.e., $h_j(i_1)\!\ne\!h_j(i_2)$ when $i_1\!\ne\!i_2$,
the MinHash sketch of $S$ can be simply represented as a vector $(\phi_1(S), \ldots, \phi_k(S))$, where each element $\phi_j(S)$ is randomly sampled with replacement from $S$ by function $h_j$.
For any two sets $S_1$ and $S_2$,
we easily find that $h_j^*(S_1\cup S_2)\!=\!\min (h_j^*(S_1), h_j^*(S_2)),  1\le j\le k.$
Therefore, the underlying MinHash sketch of the union $\phi_j(S_1\cup S_2)\!=\!\phi_j(S_1)$ when $h_j(\phi_j(S_1))\le h_j(\phi_j(S_2))$ and $\phi_j(S_2)$ otherwise.
$\phi_j(S_1\cup S_2)$ is an item in $S_1 \cap S_2$ if and only if $\phi_j(S_1)\!=\!\phi_j(S_2)$,
and we can have $P(\phi_j(S_1)\!=\!\phi_j(S_2))\!=\!P(\phi_j(S_1\cup S_2)\in (S_1\cap S_2))\!=\!\frac{|S_1\cap S_2|}{|S_1\cup S_2|}\!=\!J(S_1, S_2).$
Moreover, one can extend MinHash to handle each element $(u, i, a)$ arriving on fully dynamic stream $\Pi$ as follows: 
\textbf{case 1}) when $a\!=\!``+"$, update $\phi_j$ like a regular MinHash, i.e., set $\phi_j(S_u)\!=\!i$ if $\phi_j(S_u)\!=\!\emptyset$ or $h_j(i)< h_j(\phi_j(S_u))$ and keep $\phi_j(S_u)$ unchanged otherwise;
\textbf{case 2}) when $a\!=\!``-"$ and $\phi_j(S_u)\!=\!i$, set $\phi_j(S_u)\!=\!\emptyset$;
\textbf{case 3}) when $a\!=\!``-"$ and $\phi_j(S_u)\!=\!\emptyset$, keep $\phi_j(S_u)\!=\!\emptyset$.
However, this extension of MinHash samples an item not according to uniform distribution when item-unsubscriptions occur.
The sampling bias is not only related with the number of user's subscribed items but also the order of item subscriptions and unsubscriptions occurred in stream $\Pi$.
It is challenging to model and remove the sampling bias when a user has more than one unsubscriptions in stream $\Pi$.

To reduce the time cost,
OPH~\cite{Linips2012} only uses one hash function $h$ to process each item,
which is a random permutation from $I\!=\!\{0,1,\ldots, p-1\}$ to $I$ itself and $p$ is the maximum number of items.
OPH equally divides $I$ into $k$ bins: $[\frac{p(j-1)}{k}, \frac{pj}{k})$, $1\le j\le k$.
For a set $S \subset I$, define $S(h,j)\!=\!\{i: i\in S\wedge h(i)\in [\frac{p(j-1)}{k}, \frac{pj}{k})\}$,
and then OPH computes a variable $oph_j(S)\!=\!h^*(S(h,j))$ when $S(h,j)\ne \emptyset$ and $\emptyset$ otherwise.
At last, it estimates $J(S_1, S_2)$ as $J(S_1, S_2)\!=\!\frac{\sum_{j=1}^k \mathbf{1}(oph_j(S_1)=oph_j(S_2)\ne \emptyset)}{\sum_{j=1}^k \mathbf{1}(oph_j(S_1)\ne \emptyset \vee oph_j(S_2)\ne \emptyset)},$ and the time complexity of updating each item is $O(1)$.
Similarly, OPH can also be treated as a sampling method and exhibits a sampling bias when there exist item-unsubscriptions in $\Pi$.

Furthermore, there exist methods such as random pairing (RP)~\cite{GemullaVLDBJ2008} for uniformly sampling from dynamic graph streams.
One can extend RP to sample $k$ items $(\phi_j(S_u))_{1\le j\le k}$ (resp. $(\phi_j(S_v))_{1\le j\le k}$) from $S_u^{(t)}$ (resp. $S_v^{(t)}$).
In this case, $(\phi_j(S_u))_{1\le j\le k}$ and $(\phi_j(S_v))_{1\le j\le k}$ generated are independent,
i.e., $\phi_j(S_u)\!=\!\phi_j(S_v)$ happens with probability $\frac{1}{|S_u||S_v|}$,
which significantly differs from the probabilistic model of MinHash.
Therefore, the number of common items can be estimated as $s_{u,v}\!=\!|S_u||S_v|\sum_{j=1}^{k} \mathbf{1}(\phi_j(S_u)\!=\!\phi_j(S_v))$.
\section{Our Sketch Method} \label{sec:methods}
\begin{figure}[htb]
	\centering
	\includegraphics[width=0.45\textwidth]{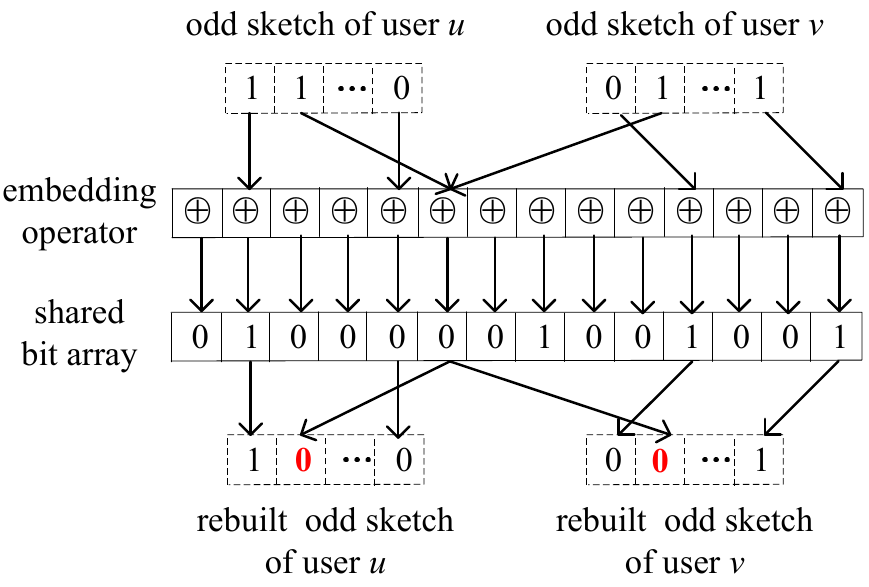}
	\caption{Overview of our method VOS. The red and bold bits are contaminated.}\label{fig:vos}
\end{figure}

Our sketch method VOS consists of a one-dimension bit array $A$ of length $m$,
a hash function $\psi$ that maps items into integers in $\{1, \ldots, k\}$ at random,
and $k$ independent hash functions $f_1, \ldots f_k$ that map users into integers in $\{1,\ldots, m\}$ at random.
As shown in Figure~\ref{fig:vos},
for each user $u$, we virtually build an odd sketch $O_u^{(t)}$ of set $S_u^{(t)}$ on the fly and embed $O_u^{(t)}$ into $A$.
Formally, $O_u^{(t)}$ is a bit array of length $k$, where each bit $O_u^{(t)}[j]$ is the parity of the number of items in $S_u^{(t)}$ of which hash value equals $j$ with respect to function $\psi$,
i.e., $O_u^{(t)}[j] = \oplus_{i\in S_u^{(t)}} \textbf{1}(\psi(i)=j), 1\le j\le k,$
where $\oplus$ is the xor operation.
The above equation tells us that any two elements $(u, i, ``+")$ and $(u, i, ``-")$ occurred before and including time $t$ offset to each other and they together are irrelevant to the value of $O_u^{(t)}$.
Our method VOS differs from the original odd sketch method~\cite{MitzenmacherWWW14} in two aspects:
1) We directly build an odd sketch $O_u^{(t)}$ of $S_u^{(t)}$ for each user $u$ rather than generate a MinHash sketch first;
2) We do not directly store the odd sketch $O_u^{(t)}$ in memory,
but use hash functions $f_1, \ldots f_k$ to randomly select $k$ bits from a shared bit array $A$ to \emph{approximately} store $O_u^{(t)}$.
In addition, for each occurred user $u$, we use a counter $n_u$ to keep tracking of the number of its subscribed items over time.
We also update bit array $A$ and a counter $\beta$ as 
$A[f_{\psi(i)}(u)] \gets A[f_{\psi(i)}(u)] \oplus 1,$
$\beta \gets \beta + \frac{2[(A[f_{\psi(i)}(u)] \oplus 1) - \frac {1}{2}]}{m},$
where $\beta$ is initialized to 0 and used to keep tracking of the fraction of 1-bits in $A$ over time,
Thus, the time complexity of updating each element is $O(1)$.

At time $t$, 
for each bit $O_u^{(t)}[j]$, $1\le j\le k$,
we randomly select a bit from $A$ using hashing function $f_j$ and xor the bit with $O_u^{(t)}[j]$, i.e.
$A[f_j(u)]\gets A[f_j(u)]\oplus O_u^{(t)}[j]$.
We easily find that the value of $A$ at time $t$ is irrelevant with the order of occurred users and  the order of the bits in their odd sketches iterated in the above procedure.
Therefore, we assume that $O_u^{(t)}[j]$ is the last bit hashed into $A$.
Let $\beta_{u,j}^{(t)}$ is the fraction of 1-bits in $A$ before the event of hashing $O_u^{(t)}[j]$ into $A$.
We easily obtain that  $O_u^{(t)}[j]$ is hashed into a 1-bit in $A$ with probability $\beta_{u,j}^{(t)}$.
Hashing $O_u^{(t)}[j]$ into $A$ changes at most one bit in $A$, therefore we have $|\beta_{u,j}^{(t)}-\beta^{(t)}|\le \frac{1}{m}$.
In this paper, we approximate $\beta_{u,j}^{(t)}$ as $\beta^{(t)}$ because $m\gg 1000$.
Based on the above observations, we model our method VOS as:
we rebuild each bit $O_u^{(t)}[j]$ at time $t$ as $\hat O_u^{(t)}[j] =  A[f_j(u)],$
which does not equal $O_u^{(t)}[j]$ with probability $P(\hat O_u^{(t)}[j] \ne O_u^{(t)}[j]) = \beta^{(t)}.$

To estimate the similarity $s_{u,v}^{(t)}$ of two users $u$ and $v$,
we first compute a sketch $\hat O_{u,v}^{(t)}$ by combining $\hat O_u^{(t)}$ and $\hat O_v^{(t)}$ using the xor operation,
i.e., $\hat O_{u,v}^{(t)}[j]\!=\!\hat O_u^{(t)}[j]\oplus \hat O_v^{(t)}[j].$
Define $\alpha_{u,v}^{(t)}$ as the fraction of 1-bits in $\hat O_{u,v}^{(t)}$, $\alpha_{u,v}^{(t)}\!=\!\frac{\sum_{j=1}^k \hat O_{u,v}^{(t)}[j]}{k},$
and $n_{u\Delta v}^{(t)}$ as the cardinality of the symmetric difference of sets $S_u^{(t)}$ and $S_v^{(t)}$,
i.e., $n_{u\Delta v}^{(t)}\!=\!|S_u^{(t)}\Delta S_v^{(t)}|\!=\!|(S_u^{(t)}-S_v^{(t)})\cup (S_v^{(t)}-S_u^{(t)})|.$
From~\cite{MitzenmacherWWW14}, we obtain $P(O_{u,v}^{(t)}[j]\!=\!1)\!=\!\frac{1-(1-2/k)^{n_{u\Delta v}^{(t)}}}{2}.$
Since $P(\hat O_u^{(t)}[j]\!\ne\!O_u^{(t)}[j])\!=\!P(\hat O_v^{(t)}[j]\!\ne\!O_v^{(t)}[j])\!=\!\beta^{(t)}$ 
we easily obtain $P(\hat O_{u,v}^{(t)}[j]\!=\!1)\!=\!((\beta^{(t)})^2 + (1-\beta^{(t)})^2)P(O_{u,v}^{(t)}[j]\!=\!1)+2\beta^{(t)}(1-\beta^{(t)}) P(O_{u,v}^{(t)}[j]\!=\!0)\!=\!\frac{1-(1-2\beta^{(t)})^2 (1-2/k)^{n_{u\Delta v}^{(t)}}}{2}.$
Then we have
\begin{equation*}\label{eq:Eaphla}
\begin{split}
\text{E}(\alpha_{u,v}^{(t)}) &= \frac{\text{E}\left(\sum_{j=1}^k \textbf{1}(\hat O_{u,v}^{(t)}[j] = 1)\right)}{k}\\
&\approx \frac{1-(1-2\beta^{(t)})^2 e^{-2n_{u\Delta v}^{(t)}/k}}{2}.
\end{split}
\end{equation*}
According to the above equation, we estimate $n_{u\Delta v}^{(t)}$ as
\[
\hat n_{u\Delta v}^{(t)} = -\frac{k(\ln (1-2\alpha_{u,v}^{(t)})- 2\ln (1-2\beta^{(t)}))}{2}.
\]
Since $s_{u,v}^{(t)} =  \frac{n_u^{(t)} + n_v^{(t)} - n_{u\Delta v}^{(t)}}{2}$, then we estimate $s_{u,v}^{(t)}$ as
\[
\hat s_{u,v}^{(t)} = \frac{n_u^{(t)} + n_v^{(t)}}{2} + \frac{k(\ln (|1-2\alpha_{u,v}^{(t)}|)- 2\ln (|1-2\beta^{(t)}|))}{4}.
\]
We easily find that the time complexity of computing $\hat s_{u,v}^{(t)}$ is $O(k)$.
Moreover, the Jaccard coefficient $\hat J(S_u^{(t)}, S_v^{(t)})$ can be estimated as $\hat J(S_u^{(t)}, S_v^{(t)})=\frac{\hat s_{u,v}^{(t)}}{n_u + n_v - \hat s_{u,v}^{(t)}}.$
Furthermore, 
the expectation and variance of its estimate $\hat s_{u,v}^{(t)}$ are computed as
\[
\text{E}(\hat s_{u,v}^{(t)}) \approx s_{u,v}^{(t)}+\frac{1}{8}-\frac{k \beta^{(t)} e^{2n_{u\Delta v}^{(t)}/k}}{(1-2\beta^{(t)})^{2}}-\frac{e^{4n_{u\Delta v}^{(t)}/k}}{8(1-2\beta^{(t)})^4},
\]
\[
\text{Var}(\hat s_{u,v}^{(t)}) \approx -\frac{k}{16}+\frac{k^2 \beta^{(t)} e^{2n_{u\Delta v}^{(t)}/k}}{2(1-2\beta^{(t)})^{2}}+\frac{k e^{4n_{u\Delta v}^{(t)}/k}}{16(1-2\beta^{(t)})^4}.
\]

\section{Evaluation} \label{sec:results}
We perform our experiments on several publicly available real-world datasets YouTube, Flickr, Orkut and LiveJournal~\cite{MisloveIMC2007}.
To generate fully dynamic graph streams including item-subscriptions and item-unsubscriptions, we follow the experiment settings in~\cite{StefaniKDD16}
and set the parameters as $q=2,000,000^{-1}$ and $d=0.5$,
which means there is a massive deletion of expected $50 \%$ edges every $2,000,000$ edges in each graph dataset.
Specially, we mainly focus on similarity estimation for users with a large number of subscribed items,
which requires extremely large memory and computational resources for similarity estimation.
Therefore, in our experiments, we first select $5,000$ users with largest cardinalities to generate user pairs of any two users in each graph dataset,
and then retain the set of user pairs that have at least one common item to keep tracking of over time.

We employ three baselines MinHash, OPH and RP as described in Section~\ref{sec:problem} and~\ref{sec:preliminaries} to compare with our method VOS for estimating similarities over time.
Here we use the metrics \emph{average absolute percentage error} (AAPE) to measure the accuracy of estimating the number of common items $\hat s_{u,v}^{(t)}$,
and \emph{average root mean square error} (ARMSE) to evaluate the performance of estimating the Jaccard coefficient similarity $\hat J(S_u^{(t)},S_v^{(t)})$ over time.
Formally, the metrics are defined respectively as $\text{AAPE}^{(t)}\!=\!\frac{1}{|P|} \sum_{(u,v) \in P} |\frac{s_{u,v}^{(t)} - \hat s_{u,v}^{(t)}}{s_{u,v}^{(t)}}|$ and $\text{ARMSE}^{(t)}\!=\!\sqrt{\frac{\sum_{(u,v) \in P}(\hat J(S_u^{(t)},S_v^{(t)}) - J(S_u^{(t)},S_v^{(t)}))^2}{|P|}}.$
In our experiments, we compare the performance of all these methods under the same memory size $m=32k|U|$ bits,
where the memory size of each value of the $k$ registers in its generated sketch for each user $u \in U$ is set as $32$ bits.
As for the parameter $k$ for the size of virtual odd sketch of our method VOS,
we can enumerate all possible values from $1$ to max-cardinality to find an optimal value for each graph dataset,
which minimizes either $\text{AAPE}^{(t)}$ or $\text{ARMSE}^{(t)}$ of selected user pairs at any time $t$.
Also we can directly set it as $\lambda$ times (i.e., $\lambda=2,3,\ldots$) larger than the memory space used by each sketch of MinHash, OPH and RP.
In the later experiments, we set $\lambda=2$.

\begin{figure}[t!]
	\centering
	\subfigure[YouTube]{\includegraphics[width=0.23\textwidth]{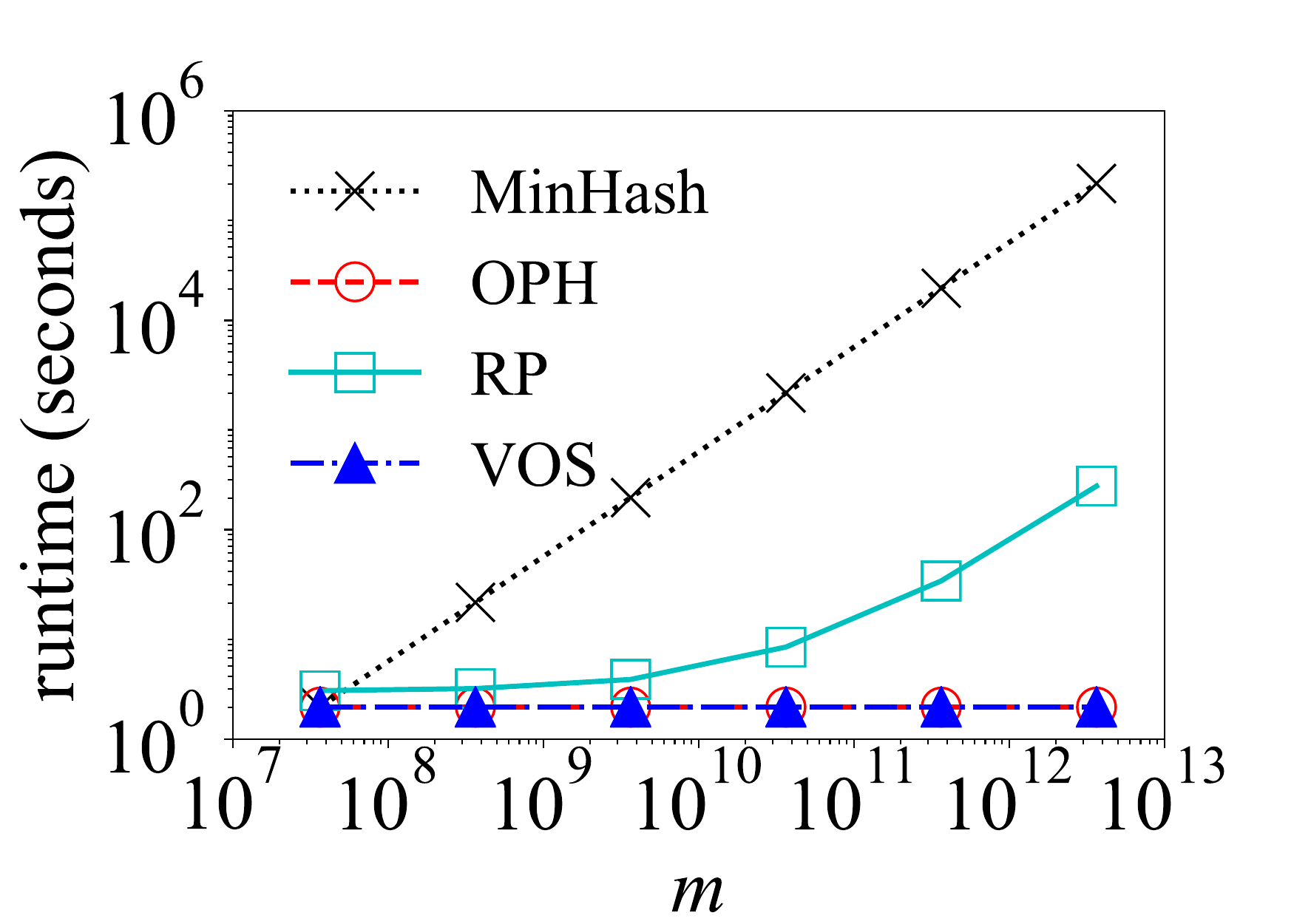}}
	\subfigure[All Datasets]{\includegraphics[width=0.23\textwidth]{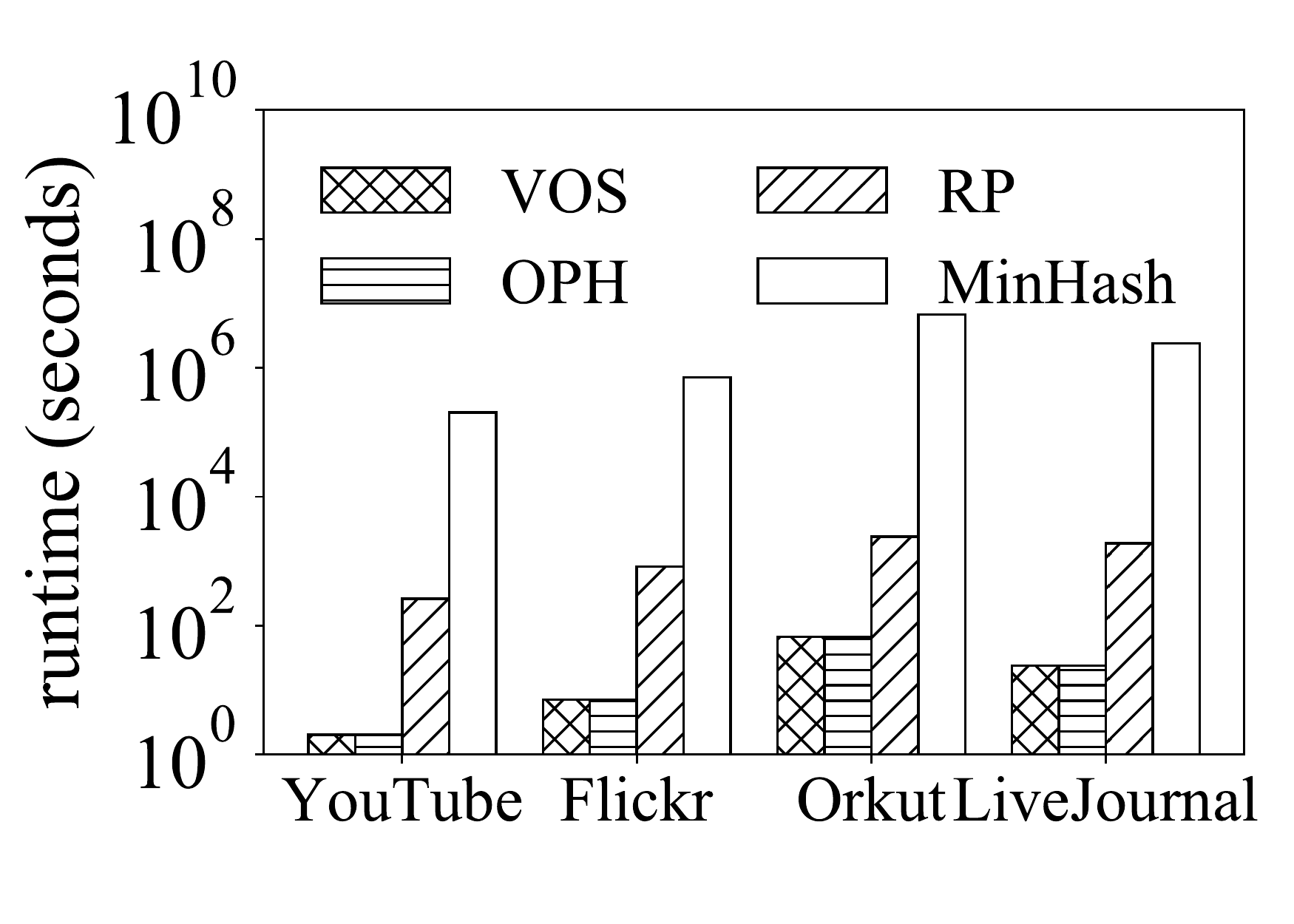}}
	\caption{Runtime of our method VOS in comparison with MinHash, OPH, and RP for different memory space $m$ (bits).}
	\label{fig:runtime}
\end{figure}
\begin{figure}[t!]
	\centering
	\subfigure[YouTube]{\includegraphics[width=0.23\textwidth]{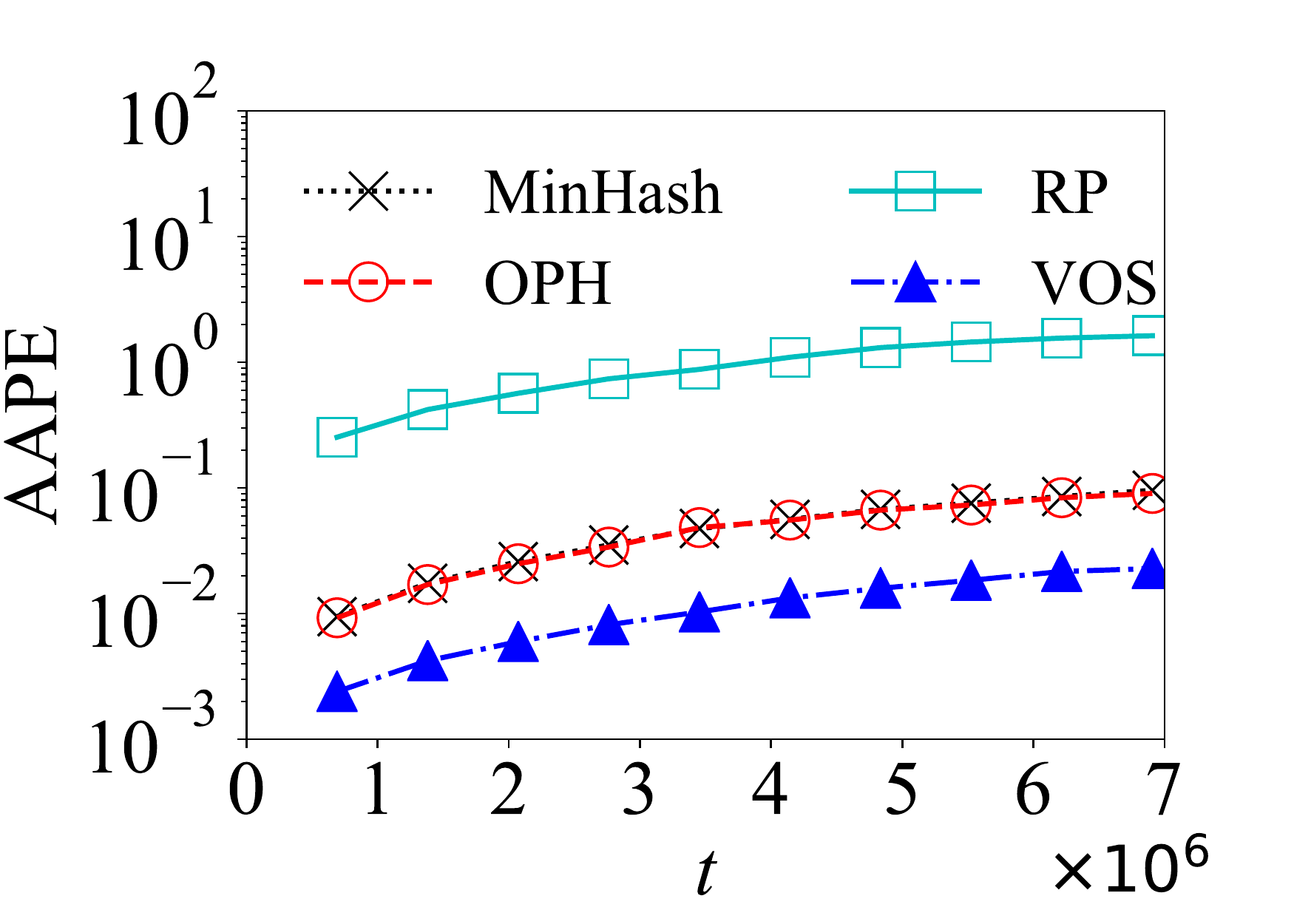}}
	\subfigure[All Datasets]{\includegraphics[width=0.23\textwidth]{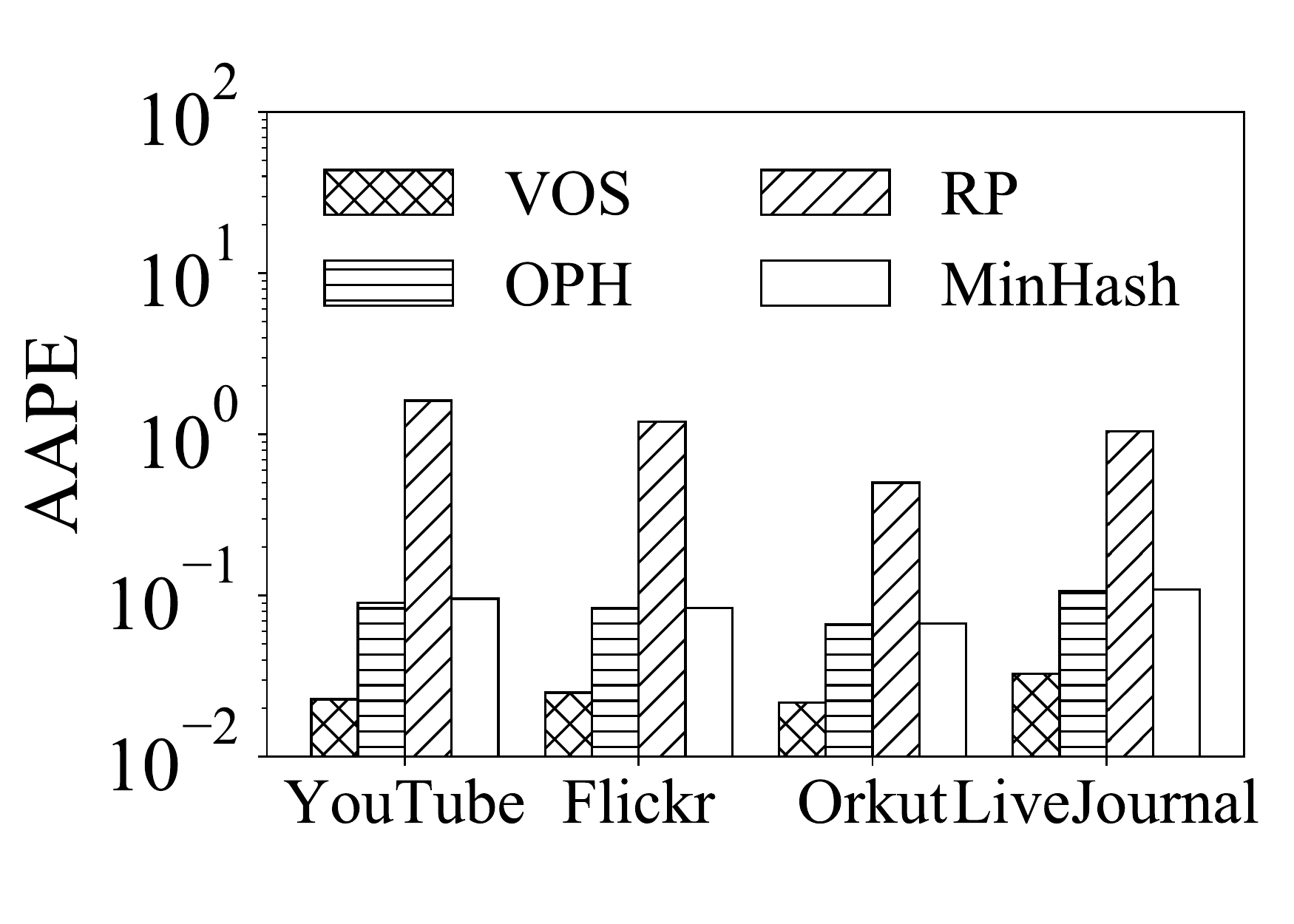}}
	\subfigure[YouTube]{\includegraphics[width=0.23\textwidth]{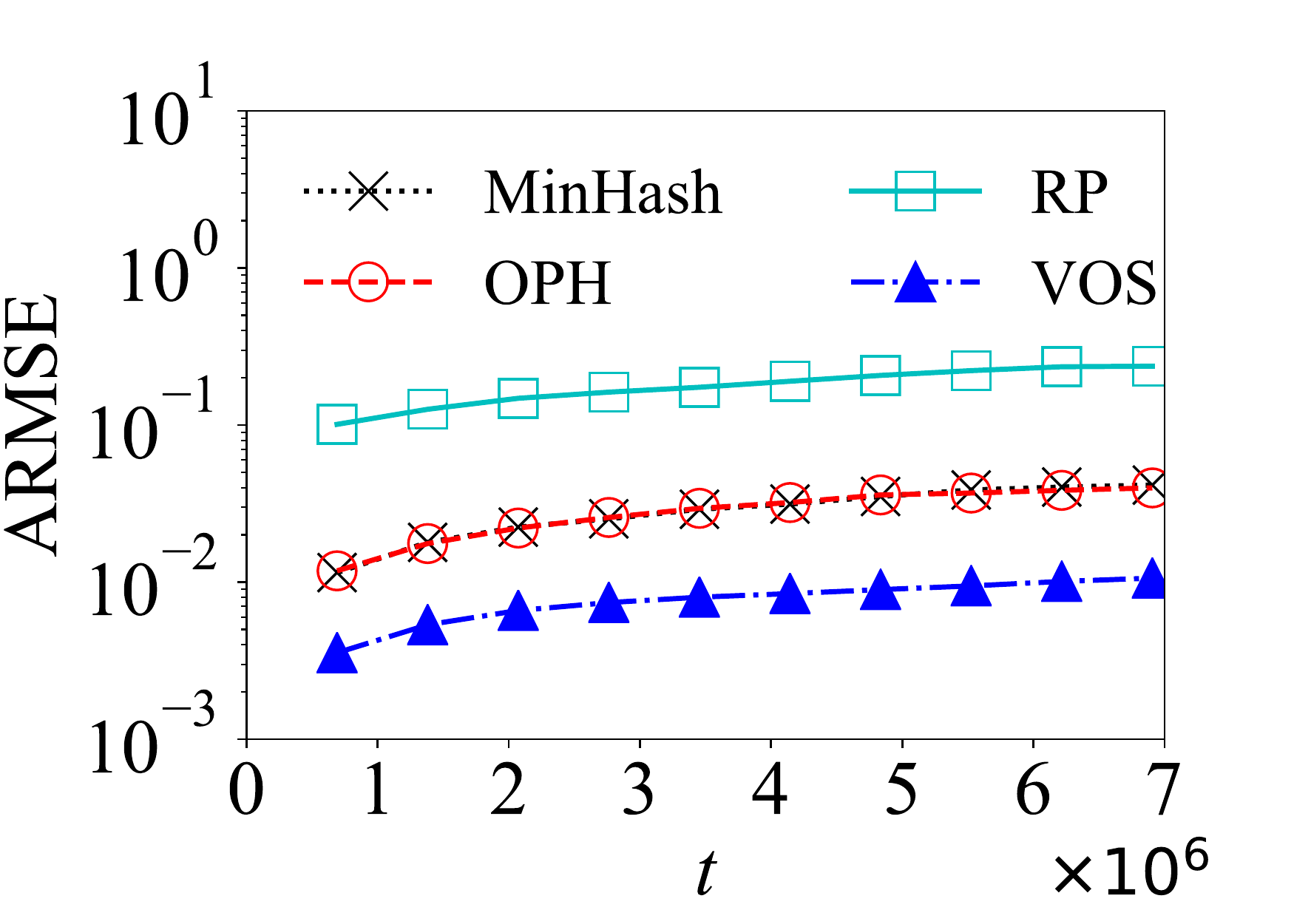}}
	\subfigure[All Datasets]{\includegraphics[width=0.23\textwidth]{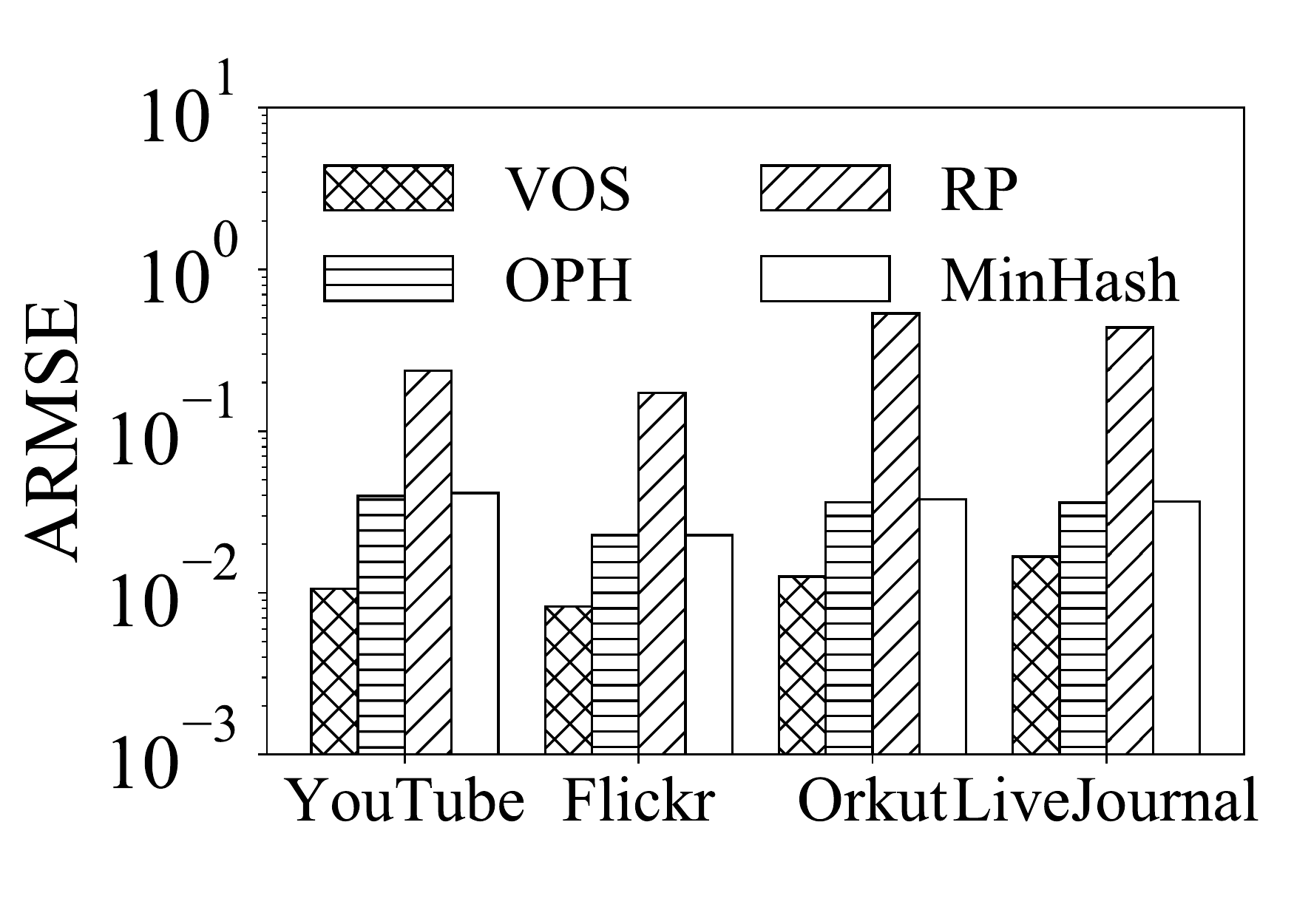}}
	\caption{Accuracy of our method VOS in comparison with MinHash, OPH, and RP for different memory space $m$ (bits) when $k=100$.}
	\label{fig:accuracy}
\end{figure}

Figure~\ref{fig:runtime} (a) shows the runtime of our method VOS in comparison with other three baselines in the dataset YouTube when the sketch size $k$ varies from $1$ to $10^5$,
and Figure~\ref{fig:runtime} (b) shows the runtime of all methods when $k=10^5$.
In our experiments, we measure the runtime during which we implement all four methods respectively to update the sketch for each user.
We can see that our method VOS and OPH are faster than MinHash and RP,
and only require the time complexity $O(1)$.
Meanwhile, we fix the sketch size $k=100$ and then compare the estimation accuracy of all four methods.
The experimental results are shown in Figure~\ref{fig:accuracy}.
Figures~\ref{fig:accuracy} (a) and (c) show the accuracy of estimating $\hat s_{u,v}^{(t)}$ and $\hat J(S_u^{(t)},S_v^{(t)})$ in dataset YouTube over time respectively,
and Figures~\ref{fig:accuracy} (b) and (d) show the accuracy in all datasets at time $t$ when all elements in graph streams arrive.
We observe that our method VOS are more accurate than other three methods and significantly reduce the estimation bias for fully dynamic graph streams.

\section{Conclusions} \label{sec:conclusions}
In this paper, we observe that state-of-the-art similarity estimation methods MinHash and OPH are indeed sampling methods for graph streams consisting of only item insertions,
and exhibit a sampling bias for fully dynamic graph streams.
To solve this problem, we develop a sampling method VOS.
VOS uniformly samples at most $k$ connected items for each user.
It fast processes each edge in the graph stream with small time complexity $O(1)$.
Based on two users' sampled items, we propose a fast method to estimate not only the Jaccard coefficient between their connected items but also the number of their common connected items.
We perform experiments on a variety of publicly available graphs,
and experimental results demonstrate that our method VOS significantly outperforms the state-of-the-art methods.


\section*{Acknowledgment}
The research presented in this paper is supported in part by National Key R\&D Program of China (2018YFC0830500), National Natural Science Foundation of China (U1736205, 61603290), Shenzhen Basic Research Grant (JCYJ20170816100819428), Natural Science Basic Research Plan in Shaanxi Province of China (2016JQ6034).

\balance
\bibliographystyle{IEEEtran}
\bibliography{COPH,randpe,ctstream}

\begin{thebibliography}{10}
\providecommand{\url}[1]{#1}
\csname url@samestyle\endcsname
\providecommand{\newblock}{\relax}
\providecommand{\bibinfo}[2]{#2}
\providecommand{\BIBentrySTDinterwordspacing}{\spaceskip=0pt\relax}
\providecommand{\BIBentryALTinterwordstretchfactor}{4}
\providecommand{\BIBentryALTinterwordspacing}{\spaceskip=\fontdimen2\font plus
\BIBentryALTinterwordstretchfactor\fontdimen3\font minus
  \fontdimen4\font\relax}
\providecommand{\BIBforeignlanguage}[2]{{%
\expandafter\ifx\csname l@#1\endcsname\relax
\typeout{** WARNING: IEEEtran.bst: No hyphenation pattern has been}%
\typeout{** loaded for the language `#1'. Using the pattern for}%
\typeout{** the default language instead.}%
\else
\language=\csname l@#1\endcsname
\fi
#2}}
\providecommand{\BIBdecl}{\relax}
\BIBdecl

\bibitem{Xia2011silo}
W.~Xia, H.~Jiang, D.~Feng, and Y.~Hua, ``Silo: A similarity-locality based
  near-exact deduplication scheme with low ram overhead and high throughput,''
  in \emph{USENIX ATC}, 2011, pp. 26--28.

\bibitem{Guo2015trustsvd}
G.~Guo, J.~Zhang, and N.~Yorke-Smith, ``Trustsvd: Collaborative filtering with
  both the explicit and implicit influence of user trust and of item ratings,''
  in \emph{AAAI}, 2015, pp. 123--129.

\bibitem{BroderSTOC2000}
A.~Z. Broder, M.~Charikar, A.~M. Frieze, and M.~Mitzenmacher, ``Min-wise
  independent permutations,'' \emph{J. Comput. Syst. Sci.}, vol.~60, no.~3, pp.
  630--659, 2000.

\bibitem{Linips2012}
P.~Li, A.~B. Owen, and C.~Zhang, ``One permutation hashing,'' in \emph{NIPS},
  2012, pp. 3122--3130.

\bibitem{ShrivastavaUAI2014}
A.~Shrivastava and P.~Li, ``Improved densification of one permutation
  hashing,'' in \emph{UAI}, 2014, pp. 732--741.

\bibitem{ShrivastavaICML2014}
------, ``Densifying one permutation hashing via rotation for fast near
  neighbor search,'' in \emph{ICML}, 2014, pp. 557--565.

\bibitem{ShrivastavaICML2017}
A.~Shrivastava, ``Optimal densification for fast and accurate minwise
  hashing,'' in \emph{ICML}, 2017, pp. 3154--3163.

\bibitem{PingWWW2010}
P.~Li and A.~C. K{\"{o}}nig, ``b-bit minwise hashing,'' in \emph{WWW}, 2010,
  pp. 671--680.

\bibitem{MitzenmacherWWW14}
M.~Mitzenmacher, R.~Pagh, and N.~Pham, ``Efficient estimation for high
  similarities using odd sketches,'' in \emph{WWW}, 2014, pp. 109--118.

\bibitem{Ioffe2010improved}
S.~Ioffe, ``Improved consistent sampling, weighted minhash and l1 sketching,''
  in \emph{ICDM}, 2010, pp. 246--255.

\bibitem{ShrivastavaNIPS2016}
A.~Shrivastava, ``Simple and efficient weighted minwise hashing,'' in
  \emph{NIPS}, 2016, pp. 1498--1506.

\bibitem{WuICDM2016}
W.~Wu, B.~Li, L.~Chen, and C.~Zhang, ``Canonical consistent weighted sampling
  for real-value weighted min-hash,'' in \emph{ICDM}, 2016, pp. 1287--1292.

\bibitem{WuWWW2017}
------, ``Consistent weighted sampling made more practical,'' in \emph{WWW},
  2017, pp. 1035--1043.

\bibitem{GemullaVLDBJ2008}
R.~Gemulla, W.~Lehner, and P.~J. Haas, ``Maintaining bounded-size sample
  synopses of evolving datasets,'' \emph{The VLDB Journal}, vol.~17, no.~2, pp.
  173--201, 2008.

\bibitem{StefaniKDD16}
L.~D. Stefani, A.~Epasto, M.~Riondato, and E.~Upfal, ``Tri{\`{e}}st: Counting
  local and global triangles in fully-dynamic streams with fixed memory size,''
  in \emph{KDD}, 2016.

\bibitem{MisloveIMC2007}
A.~Mislove, M.~Marcon, K.~P. Gummadi, P.~Druschel, and B.~Bhattacharjee,
  ``Measurement and analysis of online social networks,'' in \emph{SIGCOMM IMC}, 2007, pp. 29--42.

\end{thebibliography}
\end{document}